\documentclass[prb,twocolumn]{revtex4-1}
\usepackage[utf8x]{inputenc}
\usepackage{amsbsy}
\usepackage{graphicx}
\usepackage{color}
\usepackage{dcolumn}
\usepackage{amsmath}
\usepackage{amssymb}
\usepackage{amsfonts}
\usepackage[caption=false]{subfig}

\usepackage[colorlinks=true, citecolor=black, linkcolor=black, 
urlcolor=black]{hyperref}

\newcommand{\abinitio}{\emph{ab initio}}

\newcommand{\beq}{\begin{equation}}
\newcommand{\eeq}{\end{equation}}
\newcommand{\beqn}{\begin{eqnarray}}
\newcommand{\eeqn}{\end{eqnarray}}

\begin{document}
\title{Quantum Hydrogen-Bond Symmetrization and High-Temperature Superconductivity in Hydrogen Sulfide}

\author{Ion Errea$^{1,2}$} \email[]{ion.errea@ehu.eus}  
\author{Matteo Calandra$^{3}$} \email[]{matteo.calandra@impmc.upmc.fr} 
\author{Chris J.\ Pickard$^{4}$}
\author{Joseph Nelson$^5$}
\author{Richard J.\ Needs$^5$} 
\author{Yinwei Li$^6$}
\author{Hanyu Liu$^7$}
\author{Yunwei Zhang$^{8}$}
\author{Yanming Ma$^8$} 
\author{Francesco Mauri$^{3,9}$} \email[]{francesco.mauri@uniroma1.it} 

\affiliation{$^1$Fisika Aplikatua 1 Saila, EUITI Bilbao,
University of the Basque Country (UPV/EHU), Rafael Moreno ``Pitxitxi'' Pasealekua 3, 48013 Bilbao, 
Basque Country, Spain}
\affiliation{$^2$Donostia International Physics Center
(DIPC), Manuel Lardizabal pasealekua 4, 20018 Donostia/San
Sebasti\'an, Basque Country, Spain} 
\affiliation{$^3$IMPMC, UMR CNRS 7590, Sorbonne
Universit\'es - UPMC Univ. Paris 06, MNHN, IRD, 4 Place Jussieu,
F-75005 Paris, France}
 \affiliation{$^4$ Department of Materials Science $\&$ Metallurgy,
   University of Cambridge, 27 Charles Babbage Road, Cambridge CB3 0FS,
   UK} %cjp20@cam.ac.uk
 \affiliation{$^5$Theory of Condensed Matter Group, Cavendish
   Laboratory, J J Thomson Avenue, Cambridge CB3 0HE,
   UK} %rn11@cam.ac.uk
\affiliation{$^6$School of Physics and Electronic
Engineering, Jiangsu Normal University, Xuzhou 221116, People's
Republic of China} %yinwei_li@jsnu.edu.cn
\affiliation{$^7$ Geophysical Laboratory, Carnegie Institution of Washington, Washington D.C. 20015, USA}
% lhy@calypso.cn
\affiliation{$^8$State Key Laboratory of Superhard
Materials, Jilin University, Changchun 130012, People's Republic of
China} %mym@jlu.edu.cn
\affiliation{$^9$ Dipartimento di Fisica, Universit\`a di Roma La Sapienza, 
             Piazzale Aldo Moro 5, I-00185 Roma, Italy}

\maketitle

%%%%%%%%%%%%%%%%%% Abstract %%%%%%%%%%%%%%%%%%%%%%%%%%%%%%%%%%%%%%%%

{\bf 
Hydrogen compounds are peculiar as the quantum 
nature of the proton can crucially affect their structural and physical
properties.
A remarkable example  
are the high-pressure phases~\cite{Goncharov12071996,Loubeyre1999} of H$_2$O, where
quantum proton fluctuations favor the symmetrization of the H
bond and lower by 30 GPa the boundary between 
the asymmetric structure and the symmetric one~\cite{Benoit1998}.
Here we show that an analogous quantum symmetrization  occurs in the recently
discovered~\cite{Drozdov2015} sulfur hydride superconductor with  the record  
superconducting critical temperature $T_c=203$ K at 155 GPa. 
In this system, according to classical 
theory~\cite{Duan2014,PhysRevB.91.180502,PhysRevLett.114.157004,yinwei-unp,PhysRevB.91.060511},
superconductivity occurs via formation of a structure of
stoichiometry H$_3$S with S atoms arranged on a body-centered-cubic
(bcc) lattice. 
For $P \gtrsim 175$ GPa, the H atoms are predicted to sit
midway between two S atoms, in a structure with $Im\bar3m$
symmetry. At lower pressures the H atoms move to an
off-center position forming a short H$-$S covalent bond and a longer
H$\cdots$S hydrogen bond, in a structure with $R3m$
symmetry~\cite{Duan2014,PhysRevB.91.180502,PhysRevLett.114.157004,yinwei-unp,PhysRevB.91.060511}.
X-ray diffraction experiments confirmed the H$_3$S
stoichiometry and the S lattice sites, but were unable to
discriminate between the two phases~\cite{2015arXiv150903156E}.
Our present  \abinitio{}
density-functional theory (DFT) calculations show that the
quantum nuclear motion lowers the symmetrization pressure by 72 GPa. Consequently, we
predict that the $Im\bar3m$ phase is stable over the whole pressure
range within which a high $T_c$ was measured. The observed
pressure-dependence of $T_c$ is closely reproduced in our
calculations for the $Im\bar3m$ phase, but not for the $R3m$ phase.
Thus, the quantum nature of the proton completely rules 
the superconducting phase diagram of H$_3$S.}
%%%%%%%%%%%%%%%%%%%%%%%%%%%%%%%%%%%%%%%%%%%%%%%%%%%%%%%%%%%%%%%%%%%%

%%%%%%%%%%%%%%%%%%%%%%%%%%%%%%%%%%%%%%%%%%%%%%%%%%%%%%%%%%%%%%%%%%%%%%%%%%%%%
%
%                        FIGURES
%
\begin{figure}[t]
\includegraphics[width=0.49\linewidth]{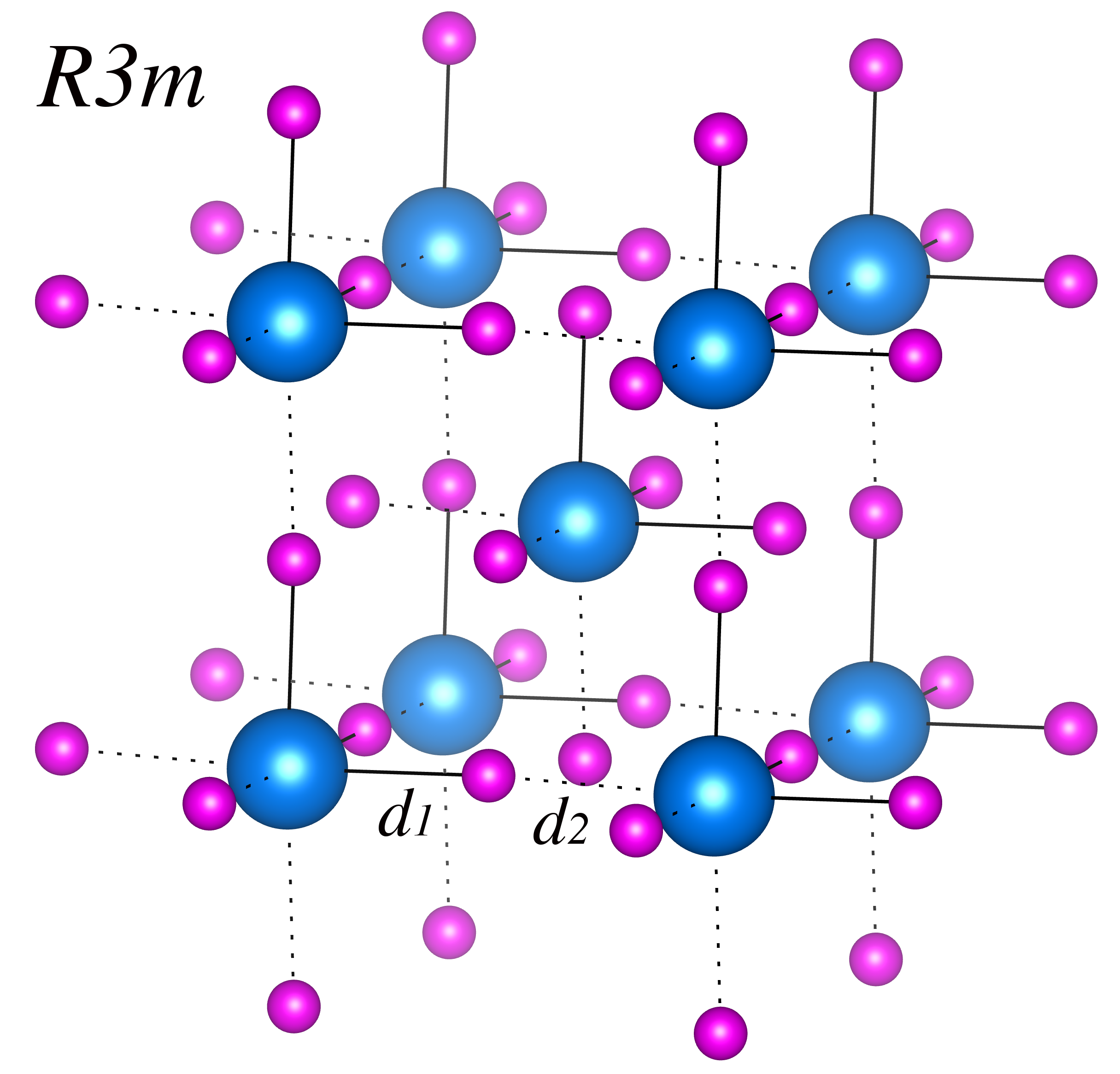} 
\includegraphics[width=0.49\linewidth]{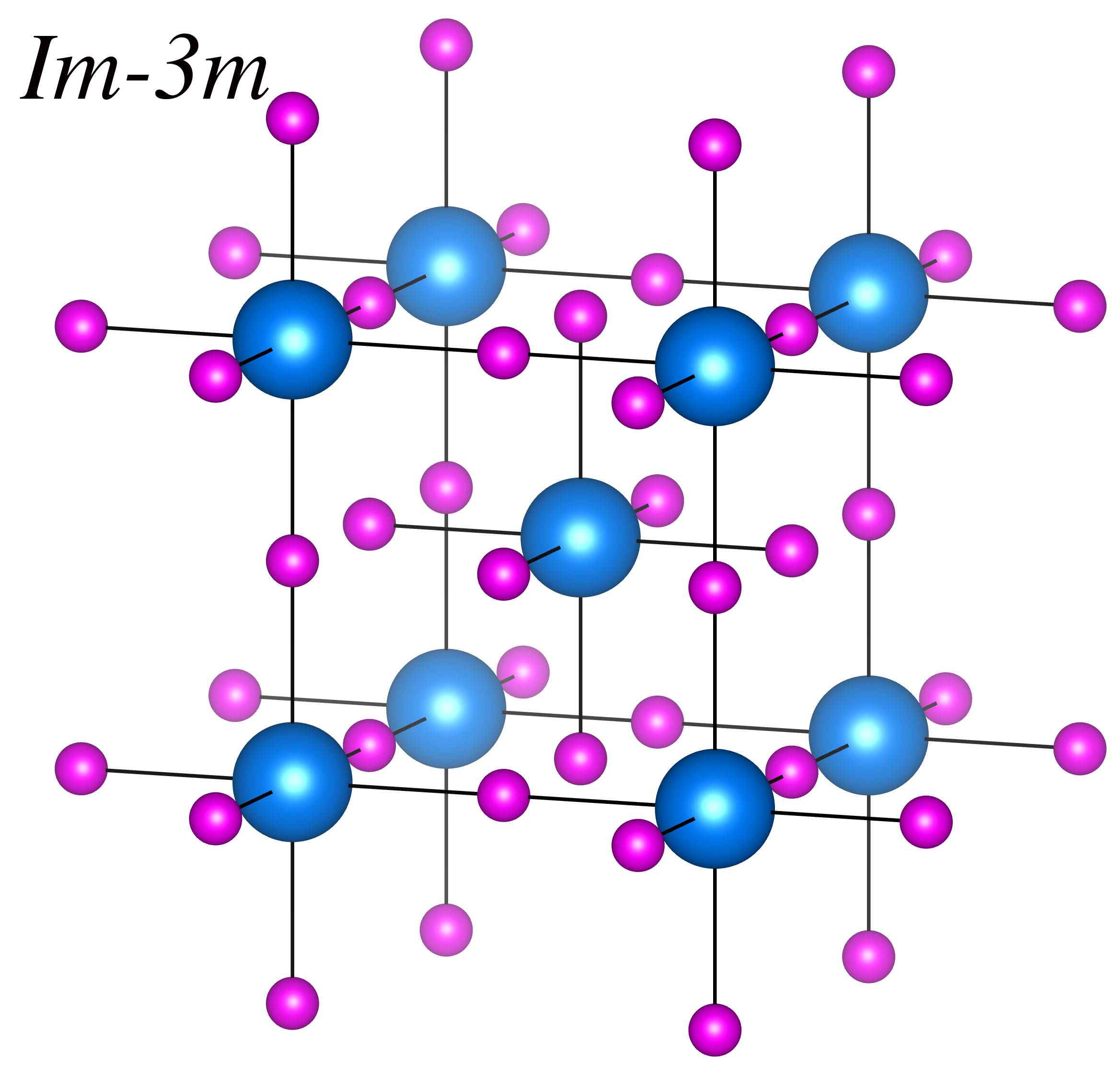} 
\caption{{\bf Crystal structures of the competing phases.} Crystal structure in the conventional bcc cell of the
         $R3m$ and $Im\bar3m$ phases. In the $R3m$ the H$-$S covalent bond of length 
         $d_1$ is marked
         with a solid line and the longer H$\cdots$S hydrogen bond of $d_2$ length
         with a dotted line. In the $Im\bar3m$ phase $d_1 = d_2$.
         } 
\label{struc}
\end{figure}

\begin{figure}[t]
\begin{center}
\includegraphics[width=0.49\linewidth]{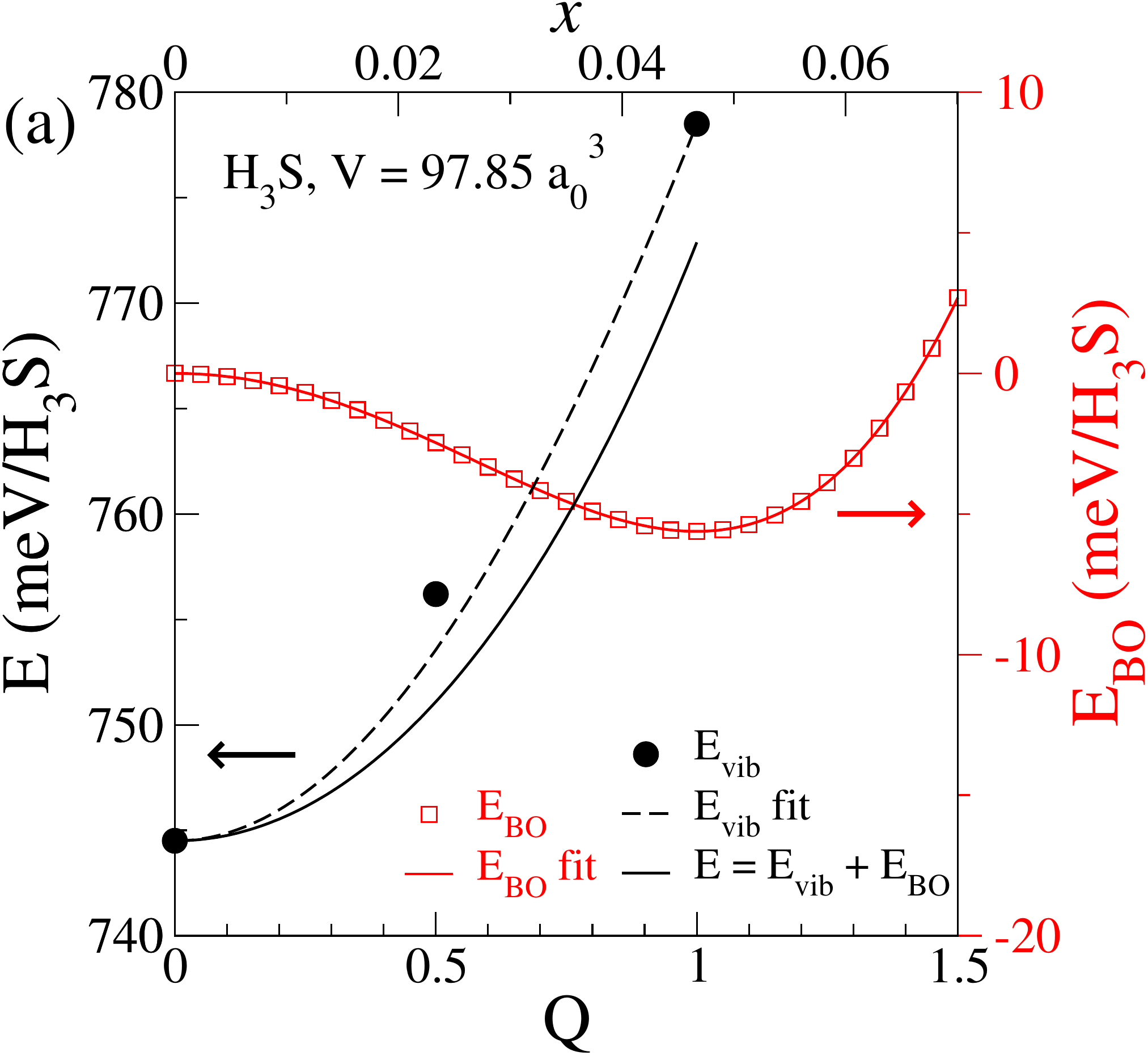}
\includegraphics[width=0.49\linewidth]{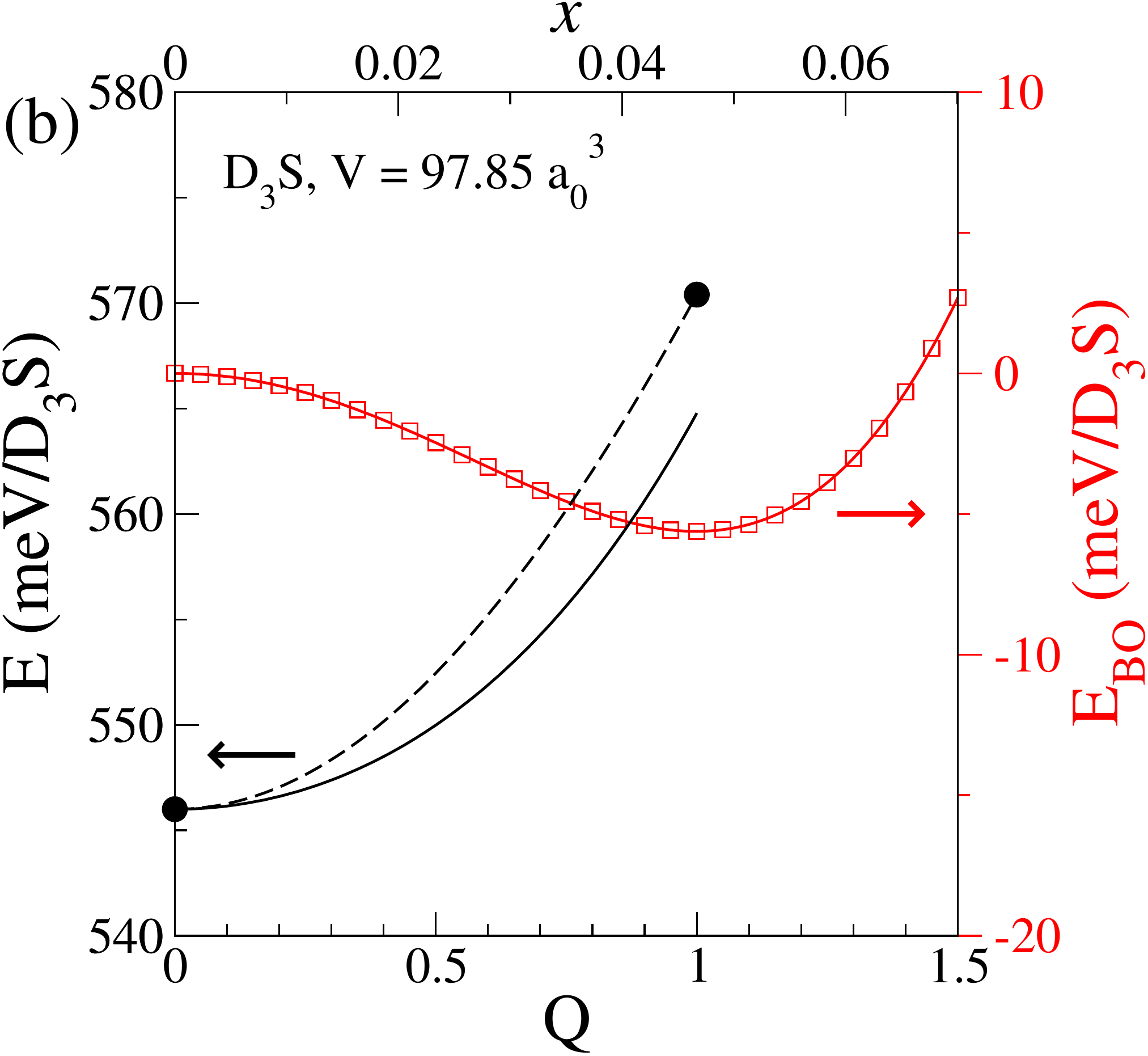} \\
\includegraphics[width=0.49\linewidth]{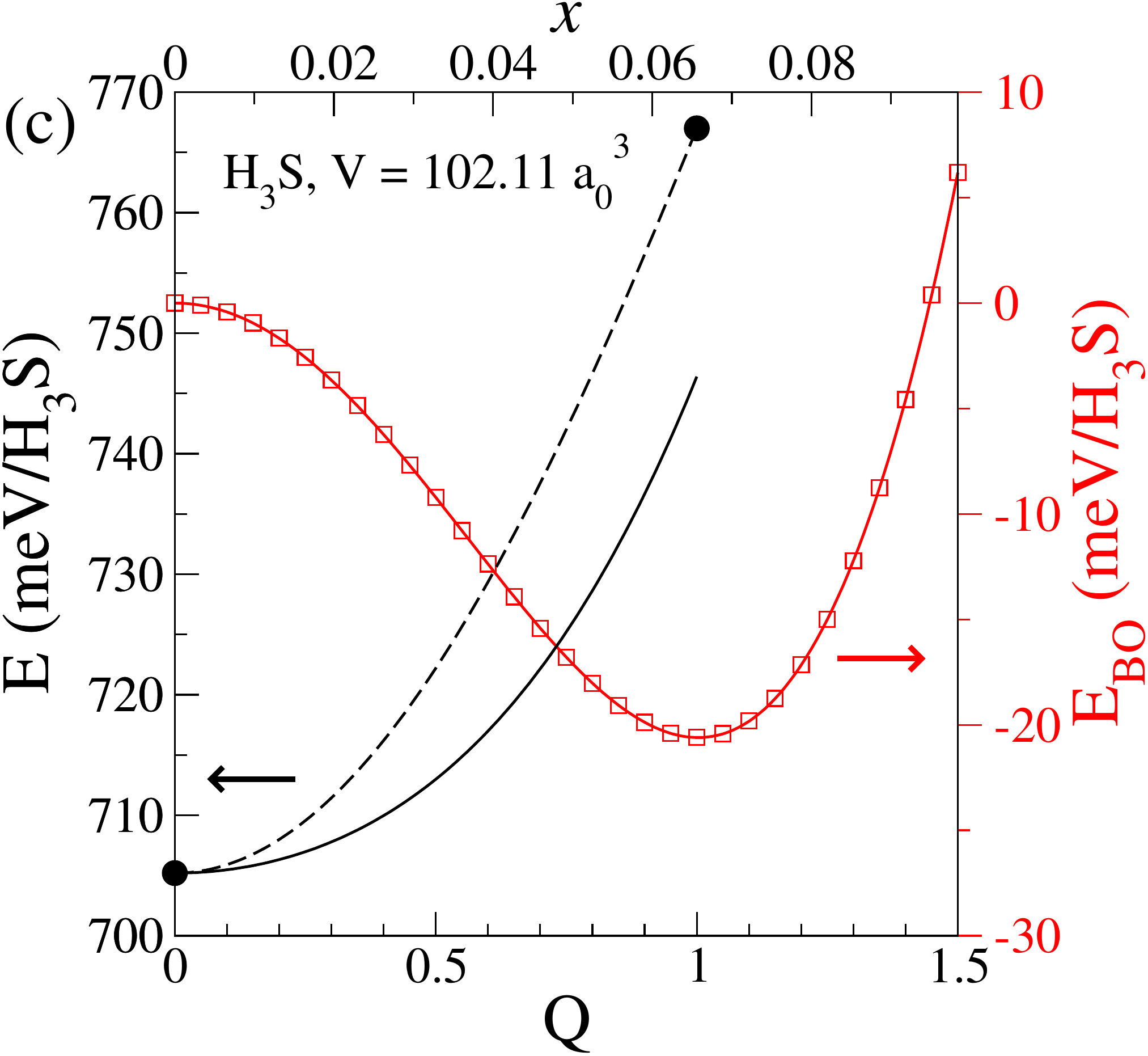}
\includegraphics[width=0.49\linewidth]{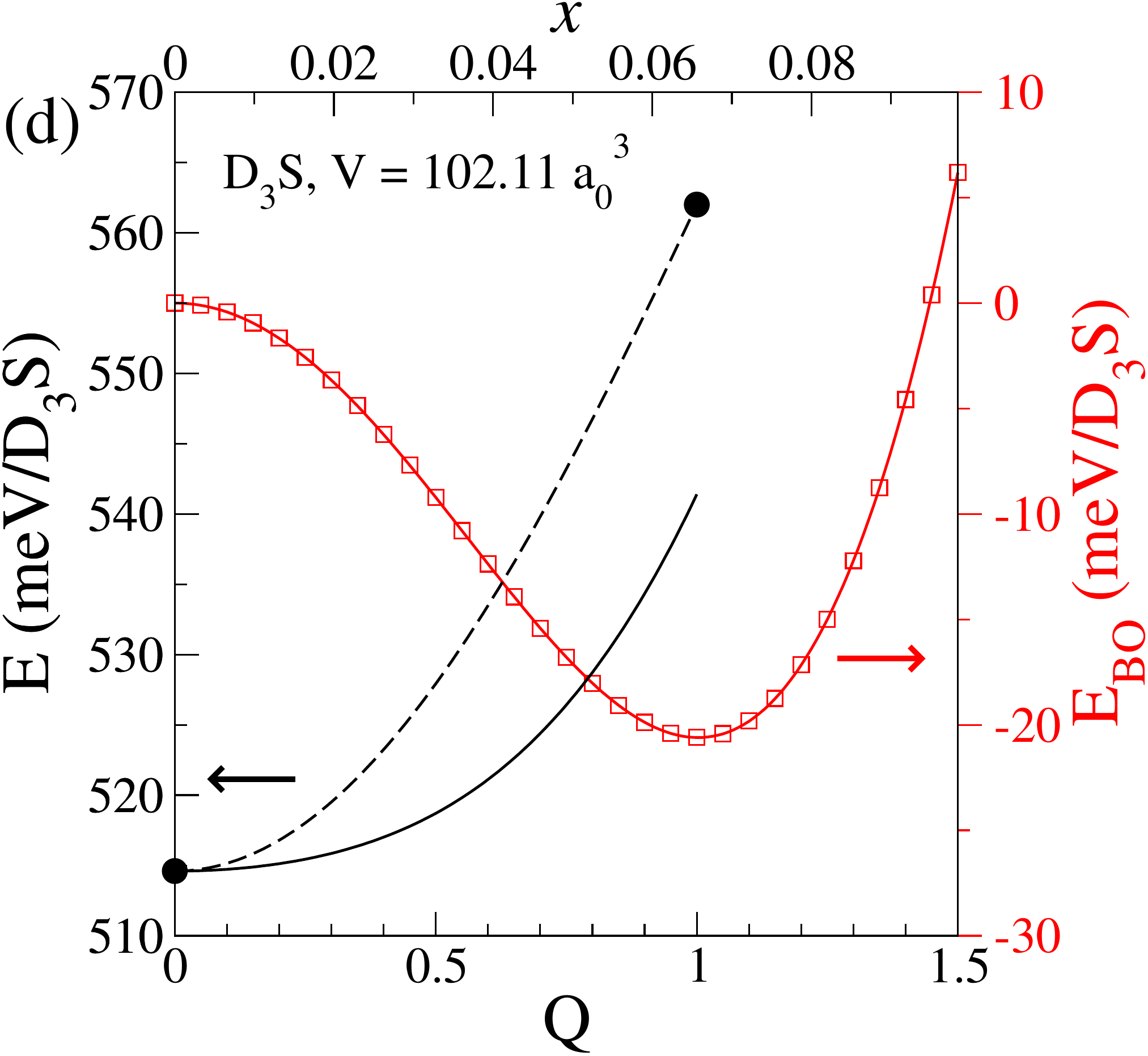} 
\end{center}
\caption{{\bf Energetics.} 
         $E_{\rm BO}$, $E_{\mathrm{vib}}$, and $E=E_{\mathrm{vib}}+E_{\rm BO}$ curves
         are shown as a function of the reaction coordinate $Q$ that transforms
         the $Im\bar3m$ structure ($Q=0$) into the $R3m$ structure 
         ($Q=1$), as well as the relative coordinate
          $x$ that measures the off-centering of the H atoms, defined as
          $x=(d_2 - a/2) / (a/2)$, where $d_2$ is the distance of the hydrogen bond and $a$ the lattice
         parameter (see Fig. \ref{struc}). 
         Note that $E_{\mathrm{vib}}$ and $E$ are plotted considering the left axis,
         while $E_{\rm BO}$ considering the right axis.
	 Crystal symmetry implies that $E(Q) = E(-Q)$, so that
	 the curves can be fitted to polynomials with only even terms.
	 This guarantees that the transition is second-order according to Landau theory~\cite{landau-vol-5}.
	 Results are presented for two different volumes of
         the primitive bcc lattice; $V = 97.85 a_0^3$ corresponds to
         approximately 150 GPa and $V = 102.11 a_0^3$ to 130 GPa.
         The pressure associated with each volume depends on both
         the isotope and $Q$. 
         Black circles represent 
         calculated $E_{\mathrm{vib}}$ points and the black dashed line
         the fitted $E_{\mathrm{vib}}(Q)$ curve (see Methods).
         The $E(Q)$ curve is obtained by addition of the fitted
         $E_{\mathrm{vib}}$ and $E_{\rm BO}$ curves.
}
\label{energ}
\end{figure}

\begin{figure}[t]
\includegraphics[width=1.00\linewidth]{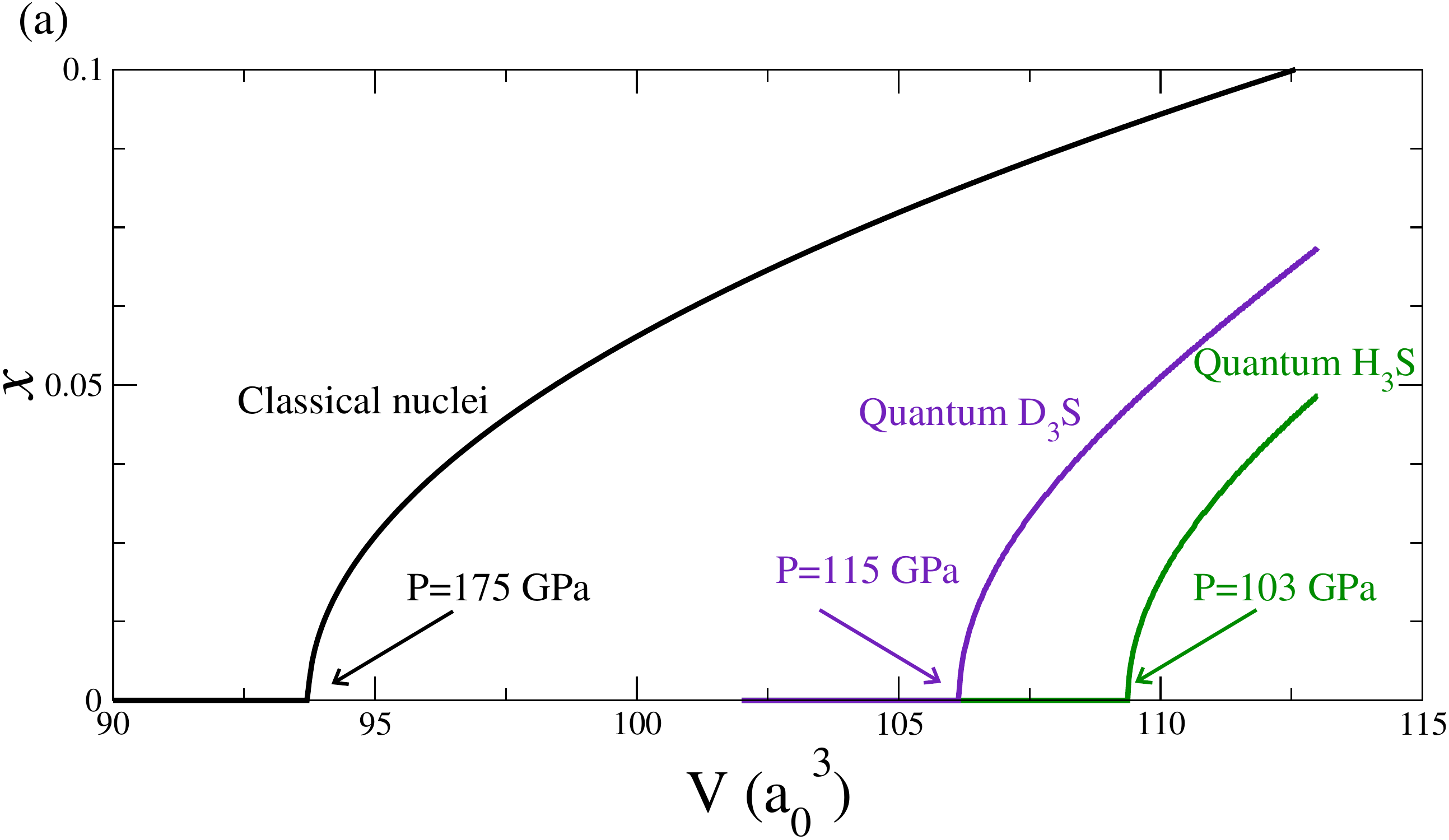} \\ 
\includegraphics[width=1.00\linewidth]{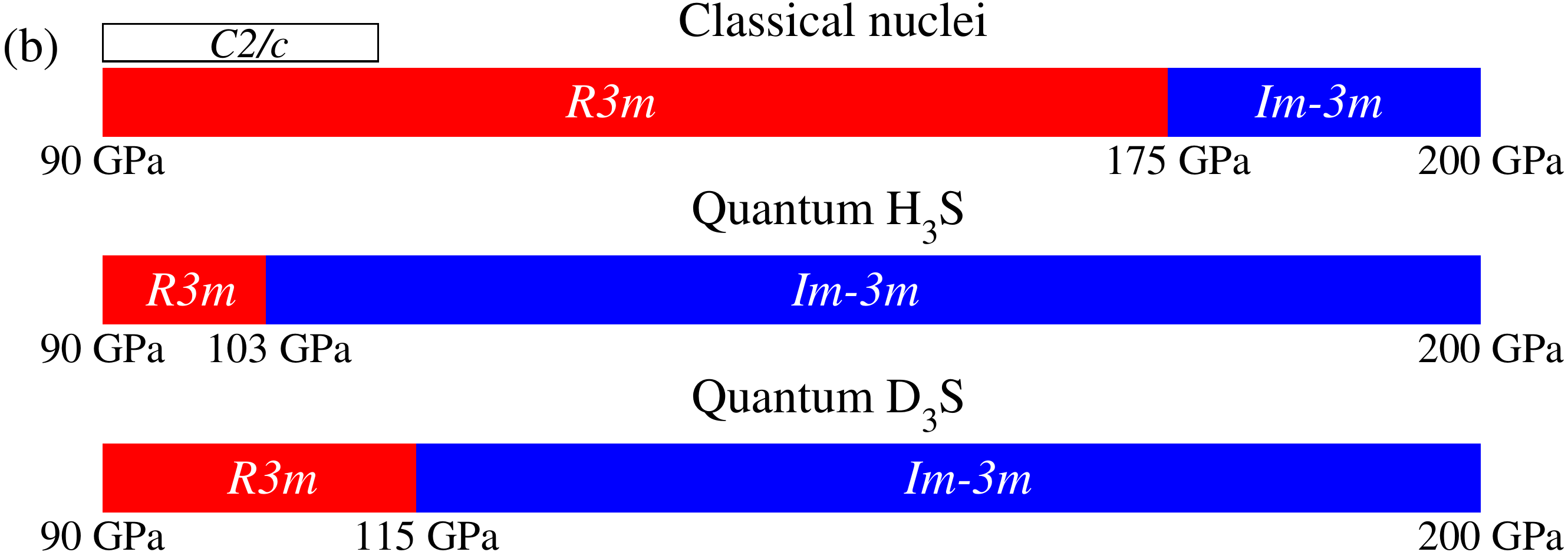} \\ 
\caption{{\bf Second-order phase transition.}
         (a) For each volume we plot the relative coordinate 
          $x$ that  yields the minimum total energy.
          $x$ measures the off-centering of the H atoms (see caption of Fig. \ref{energ}). The results are shown 
         in the classical nuclei limit, as well
         as in the quantum case both for H$_3$S and D$_3$S. The volume at which
         $x$ departs from zero marks the second-order phase transition from the
         $Im\bar3m$ phase to $R3m$ phase.
         Transition pressures are also indicated, which include  
         effects of vibrational
         energies. (b) Phase diagram for the
         second-order phase transition as a function of pressure. 
         As shown in Ref.~\onlinecite{yinwei-unp}, below 112 GPa
         H$_3$S adopts a very
         different $C2/c$ phase. We mark the expected emergence of this phase by a box.}
\label{e_c}
\end{figure}

\begin{figure*}[t]
\begin{minipage}[t]{0.4\linewidth}
  \includegraphics[width=1.00\linewidth]{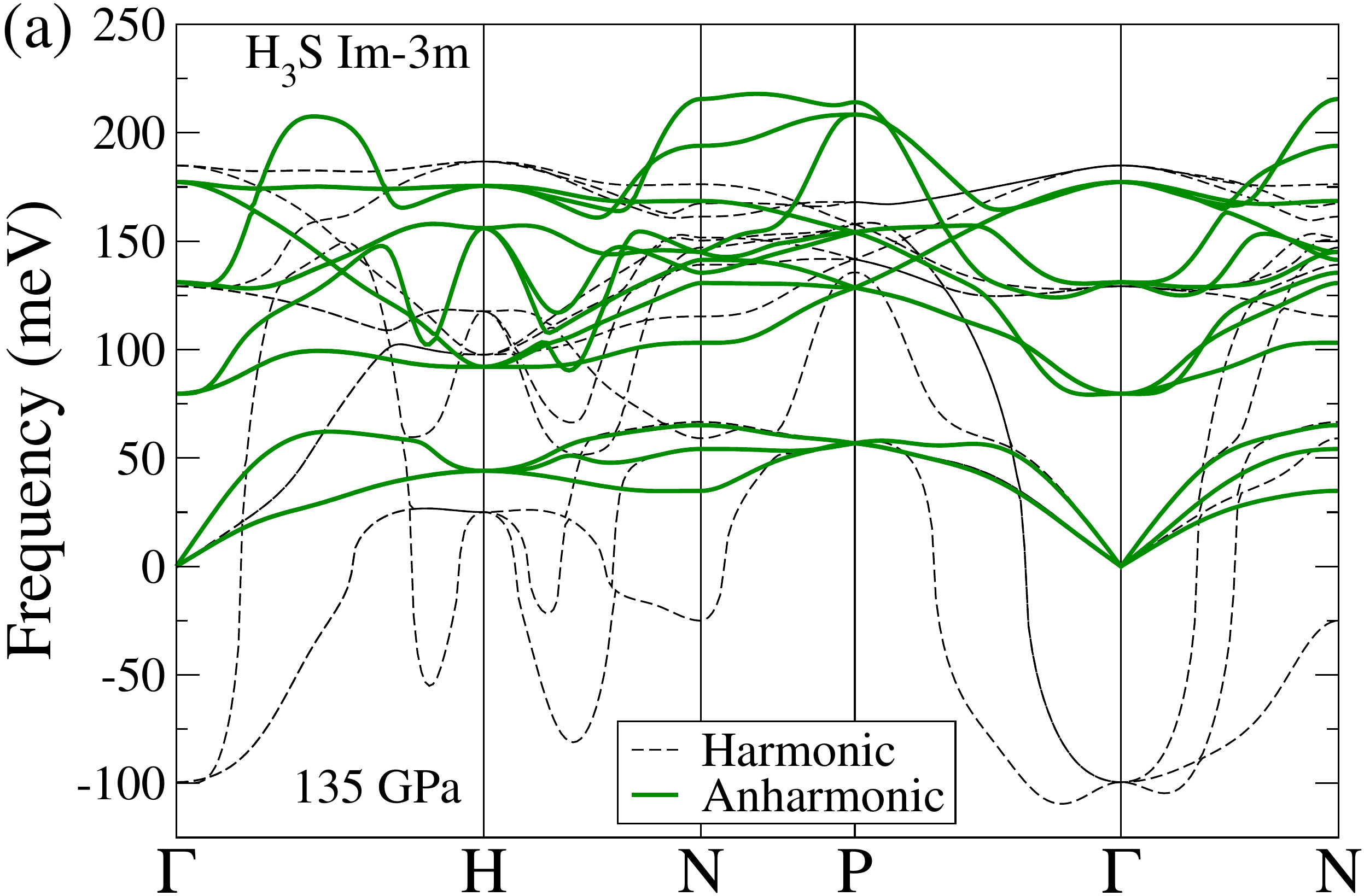} \\ 
  \includegraphics[width=1.00\linewidth]{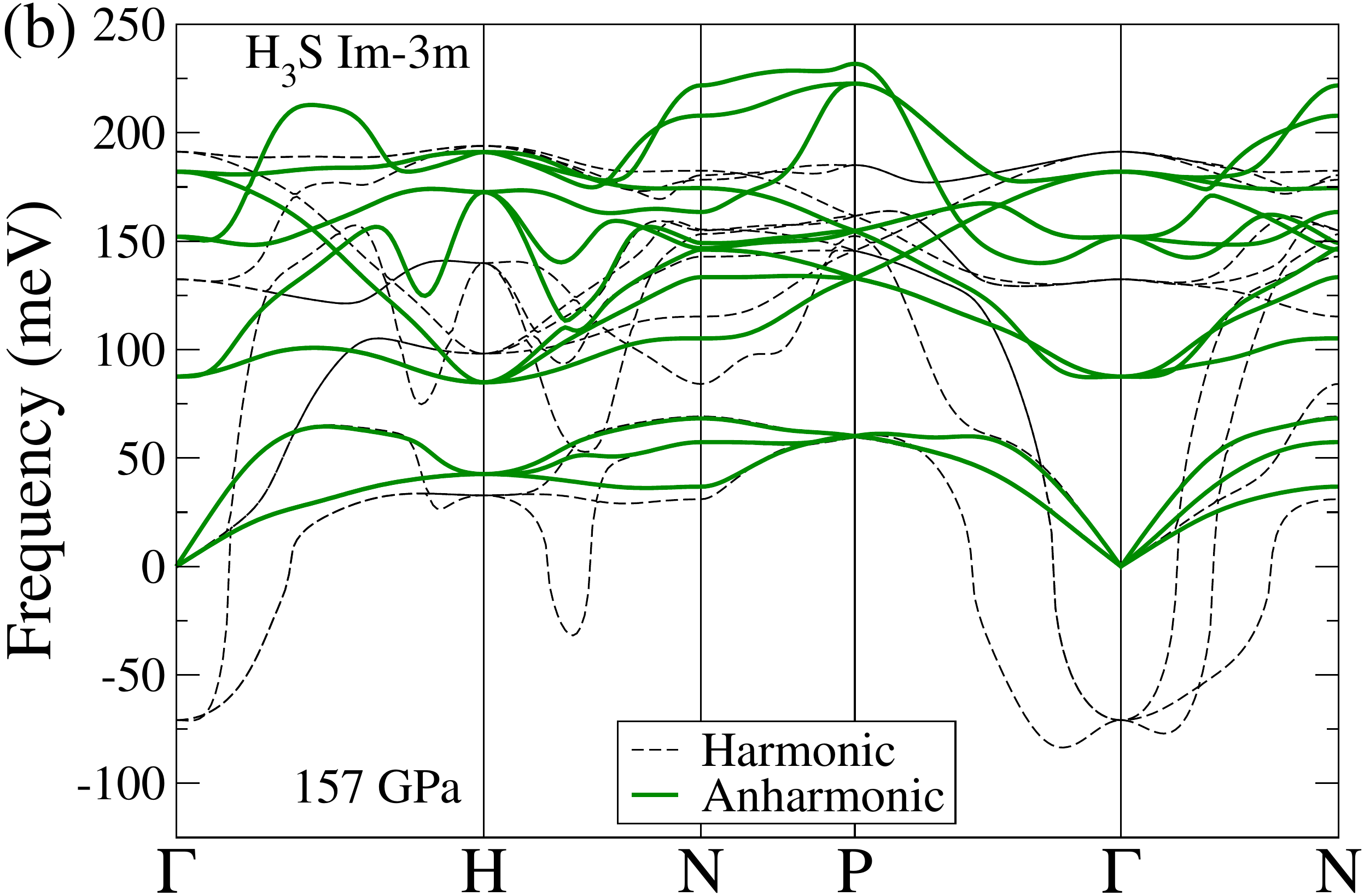} 
\end{minipage}
\begin{minipage}[t]{0.3\linewidth}
  \vspace{0.001cm}
  \vspace{-3.0cm}
  \includegraphics[width=0.80\linewidth]{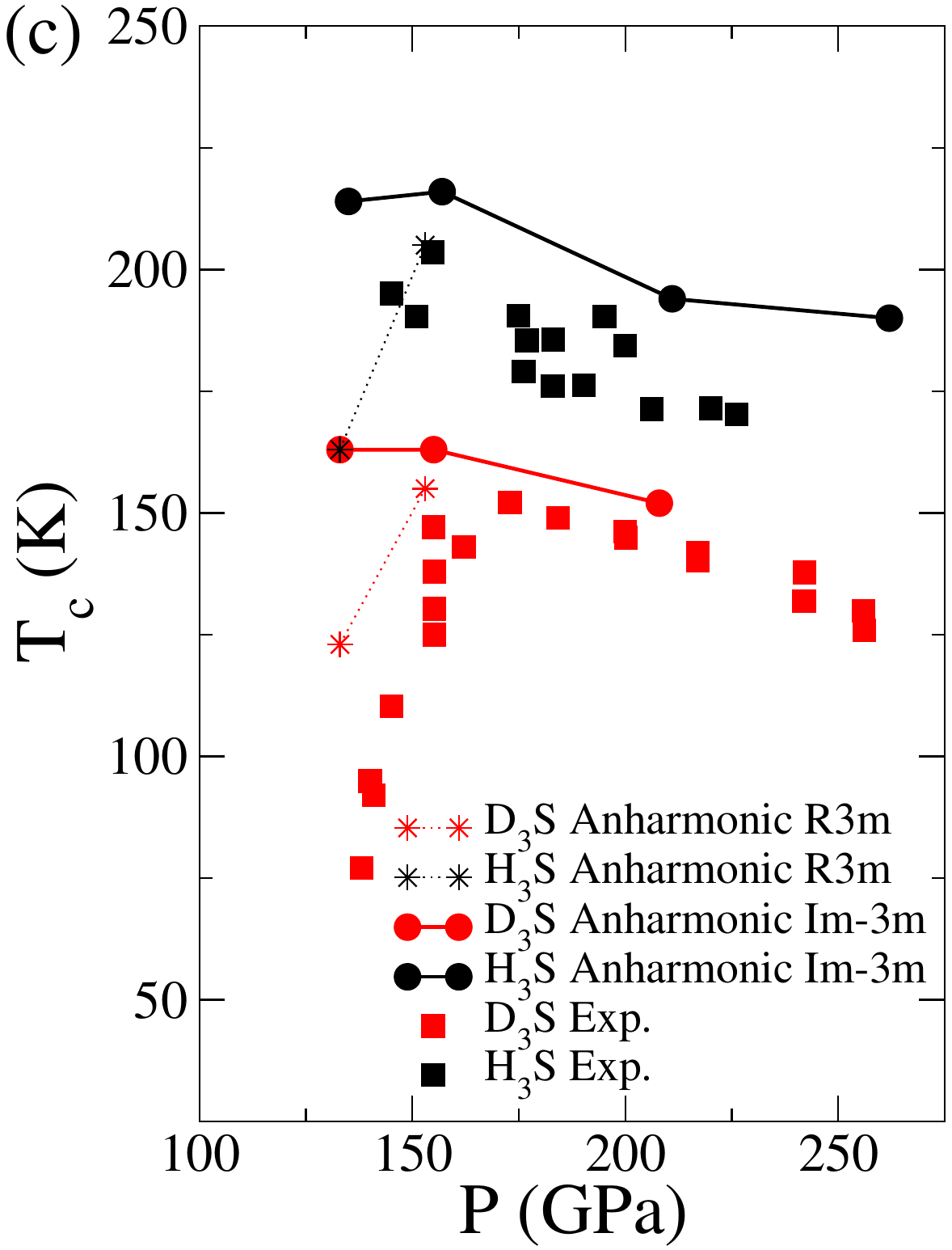}
\end{minipage}
%\begin{center}
%\includegraphics[width=0.58\linewidth]{130GPa_H3S_spectra.pdf} 
%\includegraphics[width=0.39\linewidth]{comparison_with_Eremets.pdf} \\
%\includegraphics[width=0.39\linewidth]{130GPa_D3S_spectra.pdf} \\
%\includegraphics[width=0.58\linewidth]{150GPa_H3S_spectra.pdf} 
%\includegraphics[width=0.39\linewidth]{150GPa_D3S_spectra.pdf}
%\end{center}
\caption{{\bf Phonon spectra and superconducting critical temperature.} 
         Harmonic and SSCHA anharmonic phonon spectra of the cubic high-symmetry
         $Im\bar3m$ structure for H$_3$S at different (a) 135 GPa and (b) 157 GPa.
         (c) Superconducting $T_c$'s calculated with the anharmonic phonons
         for the $Im\bar3m$ structure compared with experimental results
         obtained after annealing~\cite{Drozdov2015}. 
         $T_c$ results obtained with anharmonic phonons for the $R3m$ below 175 GPa are also shown. 
         Each pressure takes the vibrational energy into account.}  
\label{ph}
\end{figure*}

%\begin{figure}[t]
%\includegraphics[width=0.80\linewidth]{comparison_with_Eremets.pdf}
%\caption{{\bf Superconducting critical temperature.} 
%         Superconducting $T_c$'s calculated with the anharmonic phonons
%         for the $Im\bar3m$ structure compared with experimental results
%         obtained after annealing~\cite{Drozdov2015}. 
%         Results of anharmonic phonons for the $R3m$ below 175 GPa are also shown.   
%         }
%\label{tc}
%\end{figure}
%
%
%
%%%%%%%%%%%%%%%%%%%%%%%%%%%%%%%%%%%%%%%%%%%%%%%%%%%%%%%%%%%%%%%%%%%%%%%%%%%%%

The discovery of high-temperature superconductivity in compressed
hydrogen sulfide~\cite{Drozdov2015} has generated intense interest
over the last year, and has led to a number of theoretical studies
aimed at understanding the phase diagram of the H-S system as well as
the origin of the astonishingly high $T_c$
observed~\cite{Duan2014,PhysRevB.91.060511,PhysRevB.91.180502,
PhysRevB.91.184511,PhysRevLett.114.157004,PhysRevB.91.224513,PhysRevB.91.220507,
2015arXiv150106336F,yinwei-unp}. The overall consensus is that
H$_2$S, the only stable compound formed by hydrogen and sulfur at
ambient conditions, 
is metastable at high pressures and its
decomposition gives rise to several H-S compounds.
High-$T_c$ superconductivity is believed to 
occur in a structure with the H$_3$S stoichiometry, and
is considered to be conventional in nature, i.e., mediated by the electron-phonon 
interaction~\cite{Drozdov2015,Duan2014,PhysRevB.91.060511,
PhysRevB.91.184511,PhysRevLett.114.157004,PhysRevB.91.224513,PhysRevB.91.220507,
2015arXiv150106336F}.
Alternatives to conventional superconductivity
have also been discussed~\cite{Hirsch201545}.
According to structural 
predictions~\cite{Duan2014,PhysRevB.91.180502,PhysRevLett.114.157004,yinwei-unp,PhysRevB.91.060511},
H$_3$S adopts a rhombohedral $R3m$ form between approximately 112 and
175 GPa, and a cubic
$Im\bar3m$ at higher pressures.
As shown in Fig. \ref{struc},
the $R3m$ phase is characterized by covalently
bonded SH$_3$ units with a covalent H$-$S bond of 
length $d_1$. Each of these H atoms is bonded to the next S atom
by a hydrogen H$\cdots$S bond of length $d_2$. The $Im\bar3m$ phase, 
in contrast, has full cubic symmetry, with $d_1 = d_2$ so that
each H atom resides midway between the two S atoms,
as shown in Fig. \ref{struc}. 
The $R3m$ structure is nevertheless very close to
cubic symmetry, for example, the DFT-relaxed $R3m$ structure, which
represents the minimum of the Born-Oppenheimer energy surface (BOES),
has a rhombohedral angle of 109.49$^{\circ}$ at $\approx$150 GPa,
compared to 109.47$^{\circ}$ for a perfect bcc lattice. We have
verified that imposing a cubic angle on the $R3m$ structure has a
negligible effect on the energy difference between the $R3m$ and
$Im\bar3m$ structures. Consequently, we assume a cubic lattice for
both phases in the following.

The bond-symmetrizing second-order transition from $R3m$ to
$Im\bar3m$ occurs at 175 GPa according to our static lattice
calculations. At this pressure, our harmonic phonon calculations show
that a $\Gamma$-point optical phonon of the high-symmetry $Im\bar3m$
phase becomes imaginary, implying that $Im\bar3m$ is at a saddle point
of the BOES between 112 and 175 GPa, while the $R3m$ phase lies at the
minimum. Crystal symmetry guarantees that the
transition is of second-order type.
As occurs in the high-pressure ice X
phase~\cite{Benoit1998,PhysRevLett.69.462,PhysRevB.47.4863} and other
hydrogenated compounds~\cite{McMahon1990}, the quantum nature of the
proton can radically alter the pressure at which the second-order
phase transition occurs and, in the present case, can strongly affect
the stability of the $R3m$ phase below 175 GPa. Determining the
stability ranges of these phases therefore requires the inclusion of
vibrational zero-point energy (ZPE) alongside the static BOES energy.
However, the presence of imaginary phonon frequencies hinders
calculations of the ZPE, since the quasi-harmonic approximation breaks
down, and anharmonicity becomes a crucial ingredient.

To elucidate the role of anharmonicity and quantum effects in the 
pressure range in which the record $T_c$ was observed, 
we make use of the stochastic self-consistent harmonic approximation 
(SSCHA)~\cite{PhysRevLett.111.177002,PhysRevB.89.064302}.
The
variational SSCHA method was devised for calculating the free energy
and phonon spectra while fully incorporating quantum and anharmonic
effects, and it is therefore perfectly suited for our purpose.  All of
the calculations presented here are performed at 0 K. Primitive cells
for the $R3m$ and $Im\bar3m$ structures contain 4 atoms (1 S atom and
3 H atoms), and therefore a particular nuclear configuration can be
described by a 12-dimensional vector $\mathbf{R}$ containing the
atomic coordinates.  In the classical limit the ZPE is neglected and
the energy of a nuclear configuration $\mathbf{R}$ is given by the DFT
Born-Oppenheimer energy $E_{\rm BO}(\mathbf{R})$.  In the SSCHA, the
ZPE is accounted for by approximating the nuclear wave-function by a
Gaussian centered on a centroid coordinate $\mathbf{R}_{\mathrm{c}}$,
which denotes the average and most probable position of the
nuclei. For a given $\mathbf{R}_{\mathrm{c}}$, the width of the
Gaussian is obtained by a variational minimization of the expectation
value of the sum of the nuclear potential and kinetic energies.  In
the following analysis it is convenient to split the SSCHA total
energy $E(\mathbf{R}_{\mathrm{c}})$ into static and
anharmonic-vibrational-ZPE contributions:
$E(\mathbf{R}_{\mathrm{c}})=E_{\rm
  BO}(\mathbf{R}_{\mathrm{c}})+E_{\mathrm{vib}}(\mathbf{R}_{\mathrm{c}})$.

We study the energy landscape $E(\mathbf{R}_{\mathrm{c}})$
along the line defined by 
$\mathbf{R}_{\mathrm{c}}(Q) = \mathbf{R}_{Im\bar3m} + Q (\mathbf{R}_{R3m} - \mathbf{R}_{Im\bar3m})$,
where $\mathbf{R}_{Im\bar3m}$ and $\mathbf{R}_{R3m}$ are, respectively,
the coordinates corresponding to the saddle point and minimum of the BOES, representing
the two different symmetries. 
Here, $Q$ is a real number
describing the reaction coordinate, so that at $Q=0$ the centroids are
located at the atomic positions of $Im\bar3m$, and at $Q=1$ at the
atomic positions of $R3m$.
Hence, $Q$ measures the off-centering of
the hydrogen nuclear wave-function and can be associated with the
relative coordinate $x=(d_2 - a/2) / (a/2)$ that quantifies the length
of the H$\cdots$S hydrogen bond with respect to the symmetric position
($a$ is the lattice parameter).
 We analyze the curve
$E(\mathbf{R}_{\mathrm{c}}(Q))\equiv E(Q)$ for a fixed primitive bcc
unit cell.
As shown in Fig. \ref{energ}(a)
for a cell volume of $97.85 a_0^3$
the $E_{\rm BO}(Q)$ curve has a shallow double-well structure
favoring the $R3m$ structure by only 5.6 meV/H$_3$S. 
However, after adding the $E_{\mathrm{vib}}(Q)$ energy
calculated with the SSCHA, the full $E(Q)$ curve shows a clear minimum
at $Q=0$, which favors the $Im\bar3m$ structure.
At a larger volume
of 102.11 $a_0^3$, which corresponds to a pressure of around 130
GPa, the minimum of $E(Q)$ is also at $Q=0$, despite
the fact that the Born-Oppenheimer well in $E_{\rm BO}(Q)$ becomes
deeper, as shown in Fig. \ref{energ}(c).
Repeating these calculations
for D$_3$S, we find that the $Im\bar3m$ structure is the most
favorable once the ZPE has been included.  We therefore conclude that
the quantum nature of the nuclei symmetrizes the hydrogen bond and
leads to a proton wave-function centered at the atomic positions of
$Im\bar3m$ for both H$_3$S and D$_3$S.
To eliminate the possibility
that the energy minimum occurs beyond the $\mathbf{R}_{\mathrm{c}}(Q)$
line studied, we performed an unconstrained SSCHA minimization,
optimizing both the width of the Gaussians and the
$\mathbf{R}_{\mathrm{c}}$ centroid positions. The results of this
minimization show again that, within stochastic error, the centroid
position obtained corresponds to the $Im\bar3m$ structure, in which
the H$-$S covalent and H$\cdots$S hydrogen bond distances equalize,
leading to symmetric hydrogen bonds.

The difference between the vibrational energies of $R3m$ and
$Im\bar3m$ as a function of the $x$ coordinate is weakly dependent on
volume. This allows us to interpolate $E(x)$ in a
wide volume range and estimate the pressure at which the proton
wave-function shifts away from the centered position. Our calculations
show that this symmetry breaking occurs at 103 GPa in H$_3$S and 115
GPa in D$_3$S (see Fig.~\ref{e_c}). The higher transition pressure in
D$_3$S is due to weaker quantum effects. This isotope effect is
similar to the one observed in the ice VII/ice X
transition~\cite{PhysRevB.68.014106}.  Considering that below 112 GPa
the $R3m$ phase is expected to transform into a very different $C2/c$
phase consisting of isolated H$_2$S and H$_2$ molecules with H$_3$S
stoichiometry~\cite{yinwei-unp}, $R3m$-H$_3$S might not be
formed. However, D$_3$S may adopt the $R3m$ structure at pressures
below the transition to the $Im\bar3m$ phase.

The quantum proton symmetrization has an enormous impact on the phonon
spectra of H$_3$S. As mentioned earlier, and shown in Fig.~\ref{ph}, 
the phonon spectra of $Im\bar3m$-H$_3$S have several
imaginary modes in the harmonic approximation below 175 GPa. This is
analogous to ice X, which has only real positive phonon frequencies
once the classical limit predicts symmetrization of the hydrogen
bond~\cite{PhysRevLett.101.085502,Marques2009,PhysRevB.89.214101}.  On
the contrary, the corresponding anharmonic SSCHA phonon spectra show
well-behaved phonon dispersion relations with positive frequencies in
the pressure range of interest (Fig.~\ref{ph}). The anharmonic
renormalization of the phonon energies is huge, especially for the H-S
bond-stretching modes in which H atoms move towards the neighboring S
atoms, which are precisely those modes which drive the second-order
phase transition between the $Im\bar3m$ and $R3m$ phases.  Therefore,
the proximity to the second-order quantum phase transition is the
origin of the strong anharmonicity.

While the bond symmetrization in ice X occurs in an insulating system, H$_3$S
is metallic and the symmetrization strongly
affects the superconductivity. Indeed, 
the calculation of the electron-phonon coupling and the superconducting
$T_c$ lend further support to the suggestion that $Im\bar3m$-H$_3$S  
yields the record $T_c$. 
We use Wannier interpolated electron-phonon matrix elements in our 
calculations~\cite{PhysRevB.82.165111} and estimate $T_c$ solving the isotropic
Migdal-Eliashberg equations. The phonon frequencies and polarizations
that enter the electron-phonon calculations are calculated using the SSCHA.
The results obtained for the $Im\bar3m$
structure using anharmonic phonon frequencies agree well with
experimental measurements of $T_c$ for H$_3$S and D$_3$S and correctly
capture the observed increase in $T_c$ with decreasing pressure.
We also find an isotope coefficient 
$\alpha=-[{\rm ln} T_c({\rm D_3S}) - {\rm ln} T_c({\rm H_3S})] / {\rm ln}2$ 
for H$\rightarrow$D substitution of
$\alpha=0.35$ at 210 GPa and $\alpha=0.40$ at 155 GPa in good
agreement with experiment (see Fig.~\ref{ph}(c)).
The electron-phonon coupling constant $\lambda$, which scales with the
phonon frequencies as $\propto 1/\omega^2$, is enhanced with decreasing pressure
due to the overall softening of the phonon modes. This explains
the smooth decrease of $T_c$ with increasing pressure. 
Between 
approximately 130 and 150 GPa the increase in $\lambda$ is compensated
by the decrease in the average phonon frequency and $T_c$ saturates.
For the $R3m$ structure we also present
SSCHA calculations keeping the centroids at the $Q=1$
position. We find a rapid drop of $T_c$ with decreasing pressure as in previous
harmonic calculations~\cite{PhysRevB.91.224513}, 
in stark contrast to the experimental data.
Therefore, the observed high-$T_c$ superconductivity cannot be
explained by H$_3$S in the $R3m$ phase.
The sudden drop in $T_c$ measured for D$_3$S below 150 GPa~\cite{Drozdov2015},
which is not present in H$_3$S, 
could arise from the need for higher temperatures
to anneal the D$_3$S sample or could indicate the symmetry breaking
that we predicted at 115 GPa.
Indeed, the predicted transition 
pressure depends on the choice of the exchange correlation
functional.
Even if our choice of the PBE exchange-correlation
functional~\cite{PhysRevLett.77.3865} appears appropriate,
based on agreement between the experimentally observed equation of 
state~\cite{2015arXiv150903156E} and
DFT calculations, we cannot exclude a small error on the transition
pressure.

%Strong quantum and anharmonic effects, such as those described here
%for H$_3$S, might also be expected in other hydrogen-rich materials
%with prospects for high-$T_c$ superconductivity, both in terms of
%their thermodynamical stability and electron-phonon coupling strength.

%
% Methods
%

\section*{Methods}

Supercell calculations for the SSCHA~\cite{PhysRevLett.111.177002,PhysRevB.89.064302}
and linear response calculations~\cite{RevModPhys.73.515}
were performed within DFT and the generalized gradient 
approximation functional~\cite{PhysRevLett.77.3865} as implemented in the
{\sc Quantum ESPRESSO}~\cite{0953-8984-21-39-395502} code.
We used ultrasoft pseudopotentials~\cite{PhysRevB.41.7892},
a plane-wave cutoff energy of 60 Ry for the kinetic energy and 600 Ry
for the charge density. The charge density and dynamical matrices
were calculated using a 32$^3$ Monkhorst-Pack shifted
electron-momentum grid for the unit cell calculations. This mesh
was adjusted accordingly in the supercell calculations. 
The electron-phonon coupling was calculated by using electron and phonon momentum grids
composed of up to $42\times42\times42$ randomly displaced points in the Brillouin zone.
The isotropic Migdal-Eliashberg equations were solved using 512 Matsubara frequencies
and $\mu^*=0.16$.

The SSCHA calculations were performed using a
3$\times$3$\times$3 supercell for both H$_3$S and D$_3$S in the $Im\bar3m$ phase, yielding
dynamical matrices on a commensurate 3$\times$3$\times$3 $q$-point grid.  
The difference between the harmonic and anharmonic
dynamical matrices in the 3$\times$3$\times$3 phonon momentum grid was
interpolated to a 6$\times$6$\times$6 grid.  Adding the harmonic matrices to the
result, the anharmonic dynamical matrices were obtained on a $6\times6\times6$
grid.  These dynamical matrices were used for the anharmonic
electron-phonon coupling calculation.
The SSCHA calculations
for $Q=1$ were performed with a 2$\times$2$\times$2
supercell. For consistency, the vibrational energies presented in Fig. \ref{energ}
were also calculated using a 2$\times$2$\times$2 supercell. The
electron-phonon calculations for $Q=1$ were, however, performed with 
the SSCHA dynamical matrices interpolated to a 6$\times$6$\times$6 grid
from the  2$\times$2$\times$2 mesh.   

The $E_{\mathrm{vib}}(Q)$ curves in Fig. \ref{energ} 
were obtained as follows. $E_{\mathrm{vib}}$ was calculated for $Q=0$ and $Q=1$ with the
SSCHA. With the SSCHA calculation at  $Q=1$, we extracted 
$\frac{{\rm d} E_{\mathrm{vib}}}{{\rm d} Q} (Q=1)$ with no further
computational effort. Considering that the derivative of the curve at $Q=0$
vanishes by symmetry, we can get straightforwardly a 
potential of the form $E_{\mathrm{vib}}(Q) = A + BQ^2 + CQ^4$.
The $E_{\mathrm{vib}} \mathrm{fit}$ curves
presented in Fig. \ref{energ} were obtained in this way. 
The extra point obtained 
at $Q=0.5$ for H$_3$S at $V = 97.85 a_0^3$ (see Fig. \ref{energ}(a))
confirmed the validity of the fitting procedure.
The $E_{{\rm BO}} (Q)$ BOES energies were calculated for many
$Q$ points yielding an accurate fitting curve.
Fig. \ref{e_c} was obtained using a polynomial interpolation
of the BOES in the volume range shown and adding the 
$E_{\mathrm{vib}}^{R3m} - E_{\mathrm{vib}}^{Im\bar3m} (x)$
curves that are practically independent of volume.

%
% Acknowledgements   
%

\section*{Acknowledgements}

We acknowledge financial support from the Spanish Ministry of Economy
and Competitiveness (FIS2013- 48286-C2-2-P), French Agence Nationale
de la Recherche (Grant No.~ANR-13-IS10-0003-01), EPSRC (UK) (Grant
No.~EP/J017639/1), Cambridge Commonwealth Trust, National Natural
Science Foundation of China (Grants No.~11204111, 11404148, and
11274136), and 2012 Changjiang Scholars Program of China.  Computer
facilities were provided by the PRACE project AESFT and the Donostia
Internatinal Physics Center (DIPC).

%
% Authors contributions
%
%
%\section*{Author contributions}
%
%I.E., M.C., and F.M.\ performed the anharmonic and superconducting
%calculations.  All authors contributed to the design of the research
%project and to the writing of the manuscript.
%
%
%
% Additional information
%
%
%\section*{Additional information}
%
%See Supplementary Information for a symmetry analysis of the phase transition,
%calculations with different exchange-correlation functionals,
%the equations of state,
%harmonic and anharmonic phonon spectra of $Im\bar{3}m$-D$_3$S,
%anharmonic phonon spectra of the $R3m$ phase,
%and superconducting properties of both $Im\bar{3}m$ and $R3m$
%structures.
%
%The authors declare no competing financial interests.

\end{document}